\documentclass[10pt,twocolumn,letterpaper]{article}

\usepackage[pagenumbers]{cvpr} 

\usepackage{graphicx}
\usepackage{amsmath}
\usepackage{amssymb}
\usepackage{booktabs}
\usepackage{epigraph}

\usepackage[pagebackref,breaklinks,colorlinks]{hyperref}

\usepackage[capitalize]{cleveref}
\crefname{section}{Sec.}{Secs.}
\Crefname{section}{Section}{Sections}
\Crefname{table}{Table}{Tables}
\crefname{table}{Tab.}{Tabs.}

\def\cvprPaperID{2} 
\def\confName{CVPR}
\def\confYear{2024}

\begin{document}

\title{At the edge of a generative cultural precipice}

\author{Diego Porres\textsuperscript{*1,2} \qquad Alex Gomez-Villa\textsuperscript{*1,2}\\
  \textsuperscript{1}Computer Vision Center, Barcelona, Spain\\ 
  \textsuperscript{2}Universitat Aut\`onoma de Barcelona, Barcelona, Spain\\
{\tt\small \{dporres,agomezvi\}@cvc.uab.es}
}

\maketitle

\let\svthefootnote\thefootnote
\let\thefootnote\relax\footnote{*: Equal contribution}

\begin{abstract}
   Since NFTs and large generative models (such as DALLE2 and Stable Diffusion) have been publicly available, artists have seen their jobs threatened and stolen. While artists depend on sharing their art on online platforms such as Deviantart, Pixiv, and Artstation, many slowed down sharing their work or downright removed their past work therein, especially if these platforms fail to provide certain guarantees regarding the copyright of their uploaded work. Text-to-image (T2I) generative models are trained using human-produced content to better guide the style and themes they can produce. Still, if the trend continues where data found online is generated by a machine instead of a human, this will have vast repercussions in culture. Inspired by recent work in generative models, we wish to tell a cautionary tale and ask what will happen to the visual arts if generative models continue on the path to be (eventually) trained solely on generated content.
\end{abstract}

\section{Introduction}
\label{sec:intro}

\epigraph{It would be a tragedy now, at a time of incredible technological and material progress if the arts were simply to become something superficial; mere entertainment, leaving a vacuum that cannot be filled.}{Mario Vargas Llosa, 2012}

Recent advancements in generative models have yielded impressive results for generating high-resolution images. In particular interest are diffusion models \cite{diffusion} and their applications in the field of the visual arts with tools such as DALL-E \cite{dalle, dalle2}, Stable Diffusion \cite{latentdiffusion}, and Midjourney \cite{midjourney}. However, these models are not created in a vacuum: they are data-hungry algorithms trained on large-scale datasets, typically scraped from the internet, such as LAION \cite{laion400m,laion5b} and higher quality subsets such as LAION-Aesthetics \cite{laion_aesthetics}. Although not all models are transparent regarding the training data they have used, they are nonetheless dependent on data that can be found online.

However, the use and creation of these datasets carried with them the bias inherent to the authors' background and location, reflected in the generative models trained to mimic them. This is reflected in the distribution of artistic styles, artists, and periods present in it. Generative models carry the risk of amplifying these biases, which has been extensively documented  \cite{futurebias_genmodels}. This also conveys a natural reaction by the artistic community: they respond either by changing \textit{where} they share their newer artwork (conditioned on the response by the platforms they used to post before\cite{Waite.2018,JoJoesArt.2022}) to downright \textit{poisoning the data well}\cite{Leffer.2024}. We extend these points in Section \ref{sec:discussion}. 

Concretely, there currently is a problem of data variety within said datasets. This is important representation-wise, as smaller communities (both societal and artistic) are yet to be accurately represented in these large-scale datasets and, in turn, generative models. While fine-tuning the models could help tackle this issue (\eg by training a LoRA \cite{hu2022lora}), this is only the case for open-source models, and if there is available data to train with. We propose this will be harder and harder as time goes on, and as more experienced artists isolate themselves from data crawlers to protect their livelihood.

While the majority of recent work has focused on the impact that these models will have on artists \cite{aiartimpact}, we also ask how artists will naturally react and the impact that the battle for data will have on culture. Indeed, art and design are not always meant to be consumed. In the Mesoamerican culture, \textit{huipils} are rectangular fabrics worn by women, their first use dating from before the Spanish conquest. The colors and patterns in them can not only identify which group or community the wearer belongs to, but they can also depict personal and symbolical meaning \cite{huipil}, and are no strangers to being stolen or culturally appropriated \cite{huipil_appropiation, huipil_louisvuitton}. These techniques carry the risk of disappearing if the current wave of pixel-based generative models supplant the non-2D techniques required to create them.

As a final piece in our argument, to unlock the true potential of modern-day AI models, there is the usual proposal to eventually replace real data with generated or synthetic data to (re)train the models. We recall, then, that generative models trained on their outputs run the risk of failing to keep up with social change \cite{stochasticparrots}. Likewise, they exhibit \textit{knowledge collapse}, where the synthetic data tend towards an 'average' \cite{knowledgecollapse}. At its most basic example, data generated with Stable Diffusion has errors in both perspective and shadow generation of objects in the scene \cite{genmodels_projgeometry2024}. These errors are easily noticed and fixed by experienced artists, but training on outputs with wrong geometric properties only exacerbates and hence entrenches these types of issues in future models, as well as in new artists who learn from them.

There is a solution to this: Alemohammad \etal \cite{2024selfconsumingMAD} eloquently poses the training setup for recursively training generative models with synthetic data as an "autophagus" (self-consuming) loop. They coin the term Model Autophagy Disorder (MAD) and note that generative models go MAD if they are trained solely on synthetic data or if they are trained with a fixed real dataset augmented with a smaller synthetic dataset. The only solution for the generative models to not go MAD is to have available, for each generation of models, \textbf{fresh} data, which can be complemented with synthetic data. We hypothesize that we will run out of fresh real data in the future for the coming generative models (which are nowadays being trained with outputs of previous generations of generative models), in particular looking at the rate at which new data is being uploaded to some of the most popular art sharing platforms.

While the following is a work in progress, it raises alarming questions regarding the cultural end-point that these models might lead us to. If they fail to faithfully generate a particular style, exacerbate the bias associated with particular groups of people, or even fail to correctly position objects with the correct perspective, these will have serious consequences for incoming artists who are looking to use these tools for learning new skills. Lastly, if the models spit out an 'average' art style, which will it be, or is it already dominating the global culture?

\section{Data Collection}

\begin{figure*}[ht!]
     \centering
     \begin{subfigure}[b]{0.33\textwidth}
         \centering
         \includegraphics[width=\textwidth]{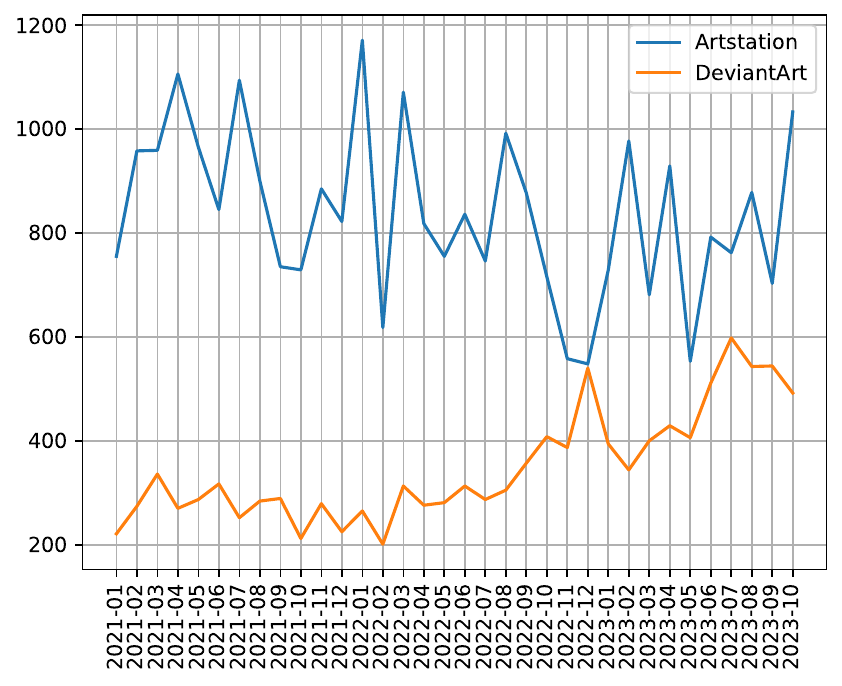}
         \caption{Total uploads}
         \label{fig:total_ubloads}
     \end{subfigure}
     \hfill
     \begin{subfigure}[b]{0.33\textwidth}
         \centering
         \includegraphics[width=\textwidth]{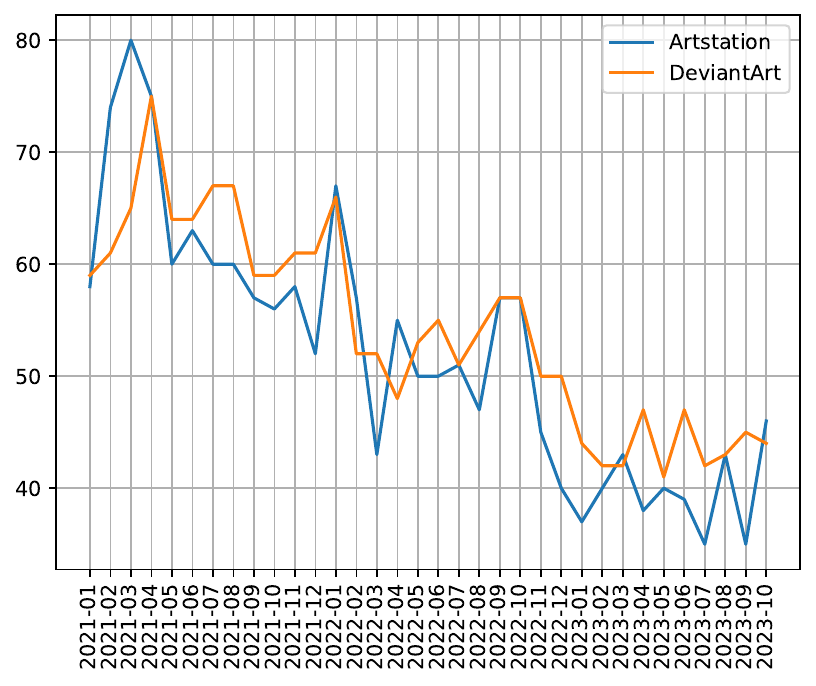}
         \caption{Uploads senior artist}
         \label{fig:total_ubloads_senior}
     \end{subfigure}
     \hfill
     \begin{subfigure}[b]{0.33\textwidth}
         \centering
         \includegraphics[width=\textwidth]{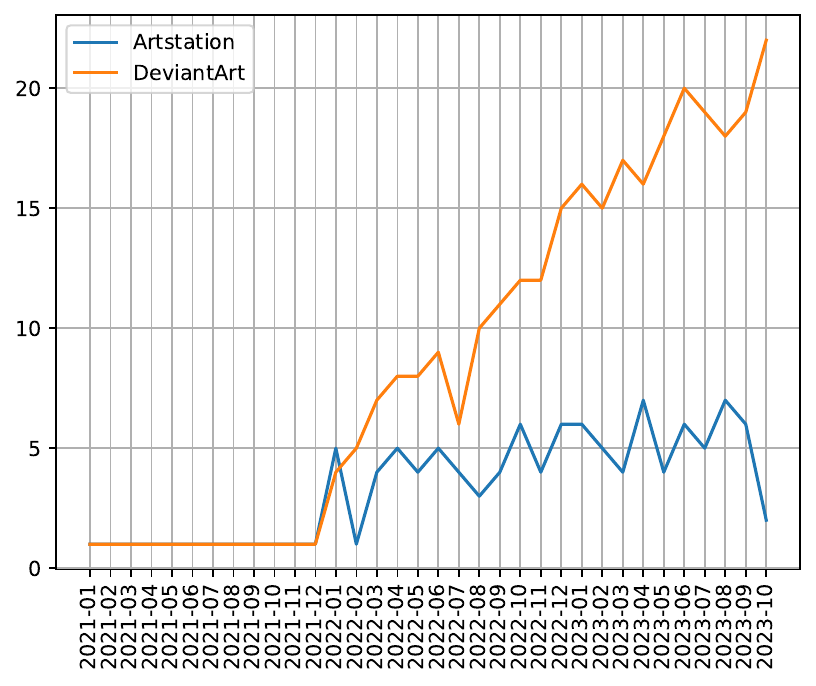}
         \caption{Uploads New artist}
         \label{fig:total_ubloads_new}
     \end{subfigure}
        \caption{Uploads Artstation and DeviantArt, sample of 250 random artists with more than 2K followers}
        \label{fig:uploads}
\end{figure*}

\addtocounter{footnote}{-1}\let\thefootnote\svthefootnote

In this section, we study the number of uploads to the popular digital art platforms Artstation\footnote{https://www.artstation.com}, DeviantArt\footnote{hhttps://www.deviantart.com/}, and Danbooru\footnote{https://danbooru.donmai.us/}. To collect data from Artstation and DeviantArt, we use the public open-source tool \texttt{gallery-dl}~\cite{gallery-dl} and randomly sample $250$ artists with more than $2000$ followers. Please note that for this work, we work with a subsample of the community due to the request limit for the web pages.

For Danbooru, we use the Danbooru2023 dataset~\cite{danbooru2023}, which is a public anime image dataset with roughly 5 million images contributed and annotated by the community. 

Analyzing the number of artists' uploads, we are looking to check trends in their behavior based on two critical events: NFTs and the release of generative models to the public (in particular of 
Stable Diffusion and DALLE-2).

Fig.~\ref{fig:total_ubloads} presents the upload from January $2021$ until October $2023$: Artstation shows a reduction in the number of uploads of artists while DeviantArt depicts an increasing trend. 

Figures ~\ref{fig:total_ubloads_senior} and ~\ref{fig:total_ubloads_new} present a different analysis utilizing the same dataset as depicted in Figure ~\ref{fig:total_ubloads}. Here, we categorize artists into seniors (those who uploaded works before January 2022) and juniors (those who only uploaded after January 2022). To ensure a balanced representation of uploads among users, we limit the count to a maximum of one publication per month per user.

Our objective of dividing the number of uploads between seniors and juniors is to examine the response to generative models that established artists have, in contrast to new artists who are already familiar with or utilize them. In Figures ~\ref{fig:total_ubloads_senior} and ~\ref{fig:total_ubloads_new}, we observe a declining trend in senior uploads and a rising trend in junior uploads.

\begin{figure}[ht!]
\includegraphics[width=0.4\textwidth]{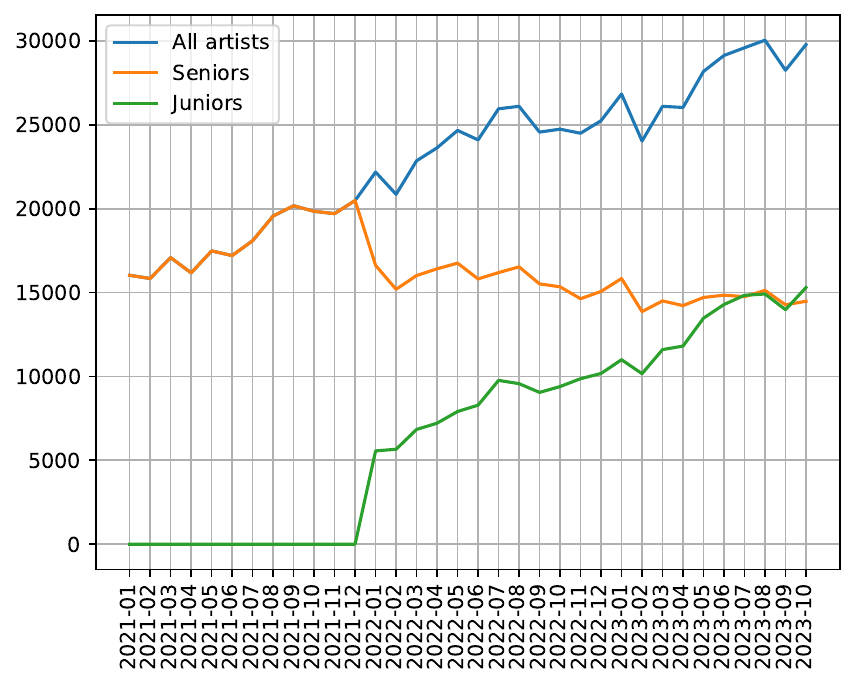}
\caption{Uploads for Danbooru2023}
\label{fig:dabooru}
\end{figure}

Fig.~\ref{fig:dabooru} shows results for the Danbooru site with the senior and junior splits. As in DeviantArt uploads, we have an increasing trend, but when we divide the artists into seniors and Juniors, a decreasing trend appears for seniors.

\section{Discussion}
\label{sec:discussion}
\subsection{Upload reduction}

If we examine users' uploads in Figs.~\ref{fig:total_ubloads} and ~\ref{fig:dabooru}, a clear trend is not evident. However, there is an apparent reduction in uploads when we focus on seniors. We believe this pattern is a consequence of the NFT explosion and further rooted by the appearance of generative model applications, as artists are concerned that their art will be used or stolen~\cite{ali2023studying,lovato2024foregrounding} without their explicit consent. While a more thorough statistical analysis should be performed here to remove confounding factors, the effect is the same, with senior artists decreasing their uploads to these easily accessible sites, giving more room to machine-generated data.  

\subsection{Technology Adoption}

Figures ~\ref{fig:total_ubloads_new} and ~\ref{fig:dabooru} illustrate that junior artists are contributing to the increasing upload trend. We believe this trend is emerging because junior artists are incorporating generative models into their creative processes, either by uploading solely generated images or by collaborating in a loop with a generative model (utilizing tools like Adobe Firefly \cite{adobefirefly}), permitting them to produce work at a much more rapid pace.

\subsection{Data poisoning}

While regulation can play a key factor in securing the data of artists as suggested in \cite{aiartimpact}, it appears that the law is not moving fast enough. Furthermore, digital art websites are releasing AI tools for their platforms, and artists who use generative models are already winning art contests \cite{pinkfloydai}. Some artists have found that data poisoning is the only way to protect their work. Methods such as watermarking (placing large logos over images) or tools like Glaze~\cite{shan2023glaze} and Nightshade~\cite{shan2023prompt} are becoming more common each day, but these are not permanent solutions.

\section{Future Work}

As mentioned in Sec. \ref{sec:discussion}, we have worked with a subsample of the whole artistic community, mainly due to computational resources. We wish to extend this work to include a wider range of artistic profiles and do a fine-grained analysis not only due to the number of followers (although it is a good filter in non-social media websites). At the same time, we must be careful when attributing the choices artists take to specific events or tool releases, for which a thorough statistical analysis is warranted. Lastly, an analysis of what each group of artists (seniors and juniors) is uploading should help clarify if specific themes or styles are affected more by these tools, or if these effects are general.

\section{Ackwnoledgements}
Diego Porres thanks the project TED2021-132802B-I00 funded by MCIN/AEI/10.13039/501100011033 and the European Union NextGenerationEU/PRTR. The authors acknowledge the support of the Generalitat de Catalunya CERCA Program and its ACCIO agency to CVC’s general activities.
{\small
\bibliographystyle{ieee_fullname}
\bibliography{PaperForReview}

\begin{thebibliography}{10}\itemsep=-1pt

\bibitem{midjourney}
{Midjourney}.
\newblock \url{https://www.midjourney.com}, July 2022.
\newblock Accessed on 04.08.2024.

\bibitem{adobefirefly}
Adobe.
\newblock {Adobe Firefly}.
\newblock \url{https://www.adobe.com/products/firefly.html}, 2023.
\newblock Accessed on 04.08.2024.

\bibitem{2024selfconsumingMAD}
Sina Alemohammad, Josue Casco-Rodriguez, Lorenzo Luzi, Ahmed~Imtiaz Humayun, Hossein Babaei, Daniel LeJeune, Ali Siahkoohi, and Richard Baraniuk.
\newblock Self-consuming generative models go {MAD}.
\newblock In {\em The Twelfth International Conference on Learning Representations}, 2024.

\bibitem{ali2023studying}
Safinah Ali and Cynthia Breazeal.
\newblock Studying artist sentiments around ai-generated artwork.
\newblock {\em arXiv preprint arXiv:2311.13725}, 2023.

\bibitem{stochasticparrots}
Emily~M. Bender, Timnit Gebru, Angelina McMillan-Major, and Shmargaret Shmitchell.
\newblock On the dangers of stochastic parrots: Can language models be too big?
\newblock In {\em Proceedings of the 2021 ACM Conference on Fairness, Accountability, and Transparency}, FAccT '21, page 610–623, New York, NY, USA, 2021. Association for Computing Machinery.

\bibitem{futurebias_genmodels}
Tianwei Chen, Yusuke Hirota, Mayu Otani, Noa Garc{\'i}a, and Yuta Nakashima.
\newblock Would deep generative models amplify bias in future models?
\newblock In {\em Proceedings of the IEEE/CVF conference on computer vision and pattern recognition}, 2024.

\bibitem{huipil}
Mexico Desconocido.
\newblock {El huipil, una prenda secular}.
\newblock \url{https://www.mexicodesconocido.com.mx/el-huipil-una-prenda-secular.html}, 2010.
\newblock Accessed on 04.08.2024.

\bibitem{diffusion}
Prafulla Dhariwal and Alex Nichol.
\newblock Diffusion models beat gans on image synthesis.
\newblock {\em CoRR}, abs/2105.05233, 2021.

\bibitem{gallery-dl}
Mike F\"ahrmann.
\newblock \texttt{gallery-dl}.
\newblock \url{https://github.com/mikf/gallery-dl}.
\newblock Accessed on 08.04.2024.

\bibitem{hu2022lora}
Edward~J Hu, yelong shen, Phillip Wallis, Zeyuan Allen-Zhu, Yuanzhi Li, Shean Wang, Lu Wang, and Weizhu Chen.
\newblock Lo{RA}: Low-rank adaptation of large language models.
\newblock In {\em International Conference on Learning Representations}, 2022.

\bibitem{aiartimpact}
Harry~H. Jiang, Lauren Brown, Jessica Cheng, Mehtab Khan, Abhishek Gupta, Deja Workman, Alex Hanna, Johnathan Flowers, and Timnit Gebru.
\newblock Ai art and its impact on artists.
\newblock In {\em Proceedings of the 2023 AAAI/ACM Conference on AI, Ethics, and Society}, AIES '23, page 363–374, New York, NY, USA, 2023. Association for Computing Machinery.

\bibitem{JoJoesArt.2022}
JoJoesArt.
\newblock No to ai generated images.
\newblock {\em Deviantart blog}.

\bibitem{Leffer.2024}
Lauren Leffer.
\newblock Artists are slipping anti-ai ‘poison’ into their art. here’s how it works.
\newblock {\em Scientific American}.

\bibitem{lovato2024foregrounding}
Juniper Lovato, Julia Zimmerman, Isabelle Smith, Peter Dodds, and Jennifer Karson.
\newblock Foregrounding artist opinions: A survey study on transparency, ownership, and fairness in ai generative art.
\newblock {\em arXiv preprint arXiv:2401.15497}, 2024.

\bibitem{knowledgecollapse}
Andrew~J. Peterson.
\newblock Ai and the problem of knowledge collapse.
\newblock {\em arXiv preprint arXiv:2404.03502}, 2024.

\bibitem{dalle2}
Aditya Ramesh, Prafulla Dhariwal, Alex Nichol, Casey Chu, and Mark Chen.
\newblock Hierarchical text-conditional image generation with clip latents, 2022.

\bibitem{dalle}
Aditya Ramesh, Mikhail Pavlov, Gabriel Goh, Scott Gray, Chelsea Voss, Alec Radford, Mark Chen, and Ilya Sutskever.
\newblock Zero-shot text-to-image generation.
\newblock {\em CoRR}, abs/2102.12092, 2021.

\bibitem{latentdiffusion}
Robin Rombach, Andreas Blattmann, Dominik Lorenz, Patrick Esser, and Bj{\"o}rn Ommer.
\newblock High-resolution image synthesis with latent diffusion models.
\newblock In {\em Proceedings of the IEEE/CVF conference on computer vision and pattern recognition}, pages 10684--10695, 2022.

\bibitem{genmodels_projgeometry2024}
Ayush Sarkar, Hanlin Mai, Amitabh Mahapatra, Svetlana Lazebnik, David Forsyth, and Anand Bhattad.
\newblock Shadows don’t lie and lines can’t bend! generative models don’t know projective geometry...for now.
\newblock In {\em Proceedings of the IEEE/CVF conference on computer vision and pattern recognition}, 2024.

\bibitem{huipil_louisvuitton}
Simon Schatzberg.
\newblock {Mexico accuses Louis Vuitton of copying indigenous design}.
\newblock \url{https://mexiconewsdaily.com/news/mexico-accuses-louis-vuitton-of-copying-indigenous-designs/}, 2019.
\newblock Accessed on 04.08.2024.

\bibitem{laion_aesthetics}
Christoph Schuhmann.
\newblock {LAION-Aesthetics}.
\newblock \url{https://laion.ai/blog/laion-aesthetics/}, Aug 2022.
\newblock Accessed on 04.08.2024.

\bibitem{laion5b}
Christoph Schuhmann, Romain Beaumont, Richard Vencu, Cade~W Gordon, Ross Wightman, Mehdi Cherti, Theo Coombes, Aarush Katta, Clayton Mullis, Mitchell Wortsman, Patrick Schramowski, Srivatsa~R Kundurthy, Katherine Crowson, Ludwig Schmidt, Robert Kaczmarczyk, and Jenia Jitsev.
\newblock {LAION}-5b: An open large-scale dataset for training next generation image-text models.
\newblock In {\em Thirty-sixth Conference on Neural Information Processing Systems Datasets and Benchmarks Track}, 2022.

\bibitem{laion400m}
Christoph Schuhmann, Richard Vencu, Romain Beaumont, Robert Kaczmarczyk, Clayton Mullis, Aarush Katta, Theo Coombes, Jenia Jitsev, and Aran Komatsuzaki.
\newblock {LAION-400M:} open dataset of clip-filtered 400 million image-text pairs.
\newblock {\em CoRR}, abs/2111.02114, 2021.

\bibitem{shan2023glaze}
Shawn Shan, Jenna Cryan, Emily Wenger, Haitao Zheng, Rana Hanocka, and Ben~Y Zhao.
\newblock Glaze: Protecting artists from style mimicry by $\{$Text-to-Image$\}$ models.
\newblock In {\em 32nd USENIX Security Symposium (USENIX Security 23)}, pages 2187--2204, 2023.

\bibitem{shan2023prompt}
Shawn Shan, Wenxin Ding, Josephine Passananti, Haitao Zheng, and Ben~Y Zhao.
\newblock Prompt-specific poisoning attacks on text-to-image generative models.
\newblock {\em arXiv preprint arXiv:2310.13828}, 2023.

\bibitem{pinkfloydai}
Wesley Stenzel.
\newblock {AI-generated entry wins Pink Floyd's Dark Side of the Moon competition}.
\newblock \url{https://ew.com/ai-wins-pink-floyd-s-dark-side-of-the-moon-video-competition-8628712}, 2024.
\newblock Accessed on 04.08.2024.

\bibitem{danbooru2023}
\textit{Nyanko}.
\newblock danbooru2023.
\newblock \url{https://huggingface.co/datasets/nyanko7/danbooru2023}.
\newblock Accessed on 08.04.2024.

\bibitem{huipil_appropiation}
Krithika Varagur.
\newblock {Mexico Prevents Indigenous Designs From Being Culturally Appropriated - Again}.
\newblock \url{https://www.huffpost.com/entry/mexico-prevents-indigenous-designs-from-being-culturally-appropriated-again_n_56e87879e4b0b25c9183afc4}, 2016.
\newblock Accessed on 04.08.2024.

\bibitem{Waite.2018}
Thom Waite.
\newblock 'i feel betrayed’: artists fight back against deviantart’s new ai project.
\newblock {\em Dazed}.

\end{thebibliography}
}

\end{document}